\documentclass[pre,aps,twocolumn]{revtex4}

\usepackage{color}
\usepackage{graphicx}
\usepackage{amsmath}
\usepackage{amssymb}
\usepackage{ifthen}
\usepackage{booktabs}

\usepackage{graphicx}
\usepackage{dcolumn}
\usepackage{bm}
\usepackage{longtable}
\usepackage{gensymb}

\newcommand{\bb}{\begin{equation}}
\newcommand{\ee}{\end{equation}}
\newcommand{\ba}{\begin{eqnarray*}}
\newcommand{\ea}{\end{eqnarray*}}
\newcommand{\rhor}{\rho({\bf r})}

\newcommand{\rr}{{\mathbf r}}
\newcommand{\dr}{{\rm d}{\bf r}}

\begin{document}

\title{Crystallisation of soft matter under confinement at interfaces and in wedges}

\author{Andrew J. Archer}
\affiliation{Department of Mathematical Sciences, Loughborough University, Loughborough LE11 3TU, UK}

\author{Alexandr Malijevsk\'y} \affiliation{{Department
of Physical Chemistry, Institute of Chemical Technology, Prague, 166 28 Praha 6, Czech Republic;} {Department of Aerosol Chemistry and Physics, ICPF, Academy of
Sciences, 16502 Prague 6, Czech Republic}}

\begin{abstract}
{The surface freezing and surface melting transitions that are exhibited by a model two-dimensional soft matter system is studied. The behaviour when confined within a wedge is also considered. The system consists of particles interacting via a soft purely repulsive pair potential. Density functional theory (DFT) is used to calculate density profiles and thermodynamic quantities. The external potential due to the confining walls is modelled via a hard-wall with an additional repulsive Yukawa potential. The surface phase behaviour depends on the range and strength of this repulsion: When the repulsion strength is weak, the wall promotes freezing at the surface of the wall. The thickness of this frozen layer grows logarithmically as the bulk liquid-solid phase coexistence is approached. Our mean-field DFT predicts that this crystalline layer at the wall must be nucleated (i.e.\ there is a free energy barrier) and its formation is necessarily a first-order transition, referred to as `prefreezing', by analogy with the prewetting transition. However, in contrast to the latter, prefreezing cannot terminate in a critical point, since the phase transition involves a change in symmetry. If the wall-fluid interaction is sufficiently long ranged and the repulsion is strong enough, surface melting can instead occur. Then the interface between the wall and the bulk crystalline solid becomes wet by the liquid phase as the chemical potential is decreased towards the value at liquid-solid coexistence. It is observed that the finite thickness fluid film at the wall has a broken translational symmetry due to its proximity to the bulk crystal and so the nucleation of the wetting film can be either first-order or continuous. Our mean-field theory predicts that for certain wall potentials there is a premelting critical point analogous to the surface critical point for the prewetting transition. When the fluid is confined within a linear wedge, this can strongly promote freezing when the opening angle of the wedge is commensurate with the crystal lattice.}
\end{abstract}

\maketitle

\section{Introduction}

Over the last three decades or so, much work has been done on wetting and related interfacial phenomena and significant progress has been made in our understanding of surface induced phase transitions -- see, e.g., Ref.\ \cite{bonn}. The development and use of both phenomenological Landau-type \cite{schick} and microscopic density functional theories (DFT) \cite{Evans79, Evans92, HM}, progress in computer simulation methods \cite{Landau:Binder} and also the advancing novel experimental techniques enabling the modification of surface structures on the molecular scale, have helped to provide deep insight into the adsorption phenomena of fluids on solid surfaces. The first theoretical studies, triggered by the seminal works of Cahn \cite{cahn} and Ebner and Saam \cite{ebner}, naturally focused on the most evident phenomena such as first-order or critical wetting, complete wetting, prewetting or layering, all taking place on a planar wall. In all of these situations a liquid-like layer (or gas-like layer, in the case of drying) that would be metastable in bulk is stabilised by the presence of the wall, which acts as a stabilising external field. Describing wetting films on the mesoscopic level, this can be understood by the concept of a liquid-gas interface being bound or repulsed by the interaction with the wall surface. To date, it can be said that our qualitative understanding of these phenomena is satisfactory, at least for simple fluid models \cite{bonn, schick, degennes85, dietrich, evans90, parry96, GBQ03, HTA14}.

Comparatively much less attention, at least from the microscopic perspective, has been devoted to the situation where the presence of an external field (i.e.\ wall) induces the phase transitions between the liquid and solid (crystalline) phases. One reason for this is the implicit expectation that most of the wall induced features exhibited by a liquid-gas system below the bulk critical temperature $T_c$, also apply for a liquid-solid system above the triple point $T_t$. After all, in Landau-type theories (i.e.\ phase field models) the two different phases are simply `labelled' by the two order parameter values $\pm 1$. Any change in symmetry at the transition is reflected in the choice of terms in the free energy functional. On this basis, much of the knowledge of the wetting transition and related phenomena obtained from the study of fluid systems can indeed be generalised to an almost arbitrary situation involving three different phases. However, the specific features of the system under consideration, particularly its symmetry, has to be taken into account. Perhaps the most striking difference between a system involving the liquid-gas and the liquid-solid coexistence is the absence of a critical point in the latter case, as the symmetry breaking leading to a transition between a fully symmetrical fluid and an anisotropic crystal cannot normally be accomplished in a continuous manner. This also implies that the symmetry between wetting and drying phenomena has no analogy in the liquid-solid surface transitions and thus surface freezing and surface melting should be strictly distinguished. Another reason relates to the fact that the liquid and solid phases are generally close in density, in contrast to the two phase at the gas-liquid transition. The density disparity between liquid and vapour phases (if not too closed to a critical point) can often be utilised in the theoretical approaches. In fact, the gas phase may in some cases be well approximated by a vacuum \cite{TMTT13}. This is clearly not the case for the crystal, which also exhibits spatial inhomogeneity that may play an important role, related to whether the crystal lattice is commensurate with the substrate and the spacing between walls in confined or finite systems. This also leads to much higher computational demands when the problem is treated numerically.

One can extend Landau order-parameter type theories by including in the theory more microscopic structural information of the crystal phase  \cite{Emmerich,Granasy}. These so-called phase-field-crystal theories can be viewed as simplified DFTs. The use of true (non-local) DFT to study adsorption phenomena involving a crystalline phase is rather sparse. This is because the crystal anisotropy requires much more demanding numerical treatment than a fluid phase. To resolve both the sharp density peaks and have a system of the size required to observe a sizeable prefreezing layer requires numerics with a space discretization having a large number of grid points. We are aware only of the ground-breaking work of Ohnesorge et al. \cite{ohnesorge1,ohnesorge2}, who studied melting of a crystal of a simple atomic system on its surface by tracking a sublimation line towards the triple point. Surface melting was also observed in the lattice DFT model in Ref.\ \cite{prestipino}. More recently, the influence of a slit confinement on crystallization of a soft repulsive model was examined \cite{likos}. For wide slits, surface melting was observed when the confining walls are purely repulsive and surface freezing was observed when the walls are attractive. We show here for a slightly different system that actually both effects can be seen at repulsive walls. An alternative to DFT is to use computer simulations and indeed various aspects of wall induced freezing and melting has been studied previously in this manner \cite{swoll, heni, auer, dijk, esztermann, LD07, sear2}.

Here, we present DFT results for a system exhibiting surface freezing and a surface melting induced by a purely repulsive wall. We consider a particularly simple two-dimensional (2D) model fluid consisting of particles interacting via a soft purely repulsive pair potential that is both finite and bounded: $v(r)=\varepsilon e^{-(r/R)^n}$, with $\varepsilon<\infty$. This is the so-called generalised exponential model with exponent $n$, or `GEM-$n$' potential. When the exponent $n=2$, this corresponds to the Gaussian core model \cite{Stillinger76, BLHM01, LLWL00, Likos01}. Here, we focus on the case $n=4$. The model should be considered as a coarse grained potential for soft particles in solution, such as polymers whose globular shape is given by their architecture, e.g.\ branched polymers such as dendrimers or star polymers \cite{Likos01, DaHa94, likos:prl:98, LBHM00, BLHM01, JDLFL01, Dzubiella_2001, LBFKMH02, likos:harreis:02, GHL04, MFKN05, Likos06, LBLM12}. We consider this model because the bulk structure and phase behaviour is well understood in both 2D and 3D \cite{MGKNL06, MGKNL07, MoLi07, LMGK07, MCLFK08, TMAL09, NKL12, ZCM10, ZC12, WS13, WS14,PS14,AWTK14}. The particle-wall interaction is also assumed to be repulsive and we model it by the Yukawa potential \cite{LBHM00, BLHM01, JDLFL01, Dzubiella_2001, LBFKMH02}. The main reason we consider the system in 2D, rather than in 3D, is because for the system sizes required to fully resolve a thick adsorbed film on a surface, the computations required are currently prohibitive in 3D. However, many of the qualitative predictions from the work presented here are relevant to 3D.

The aim of our work is a twofold: First, we seek to determine the full interfacial phase diagram comprising both prefreezing and premelting and investigate the (a)symmetry of the two transitions and also to compare these phenomena with those involving a fluid-fluid (i.e., liquid-gas or liquid-liquid) interface. In particular, we investigate how varying the parameters that determine the strength and range of the potential due to the wall influences the adsorption at the wall and the interfacial phase behaviour. The pertinent questions we seek to address are: What is the nature of the prefreezing/premelting transition? Do the phase diagrams contain any (surface) critical points? How does the interfacial phase behaviour depend on the molecular parameters? Second, we investigate the effect of the substrate geometry, in particular, we consider wedge-like confinement with a varying opening angle to examine how the value of the angle alters the freezing at the surface inside the wedge.

{The system is treated} using DFT \cite{Evans79, Evans92, HM}, which is a theory that provides a link between the observable phenomena, such as the growth of a liquid-like or a crystal-like layer at the wall and nature of the phase transitions, with the microscopic properties of the considered system. Thus, using DFT one can achieve a full description of the system behaviour directly from the first principles, i.e, merely from the knowledge of the mutual interaction between the particles  and the interaction of the particles with the wall. The appealing feature of {the GEM-$n$ models} is that a very simple approximate DFT, namely that which generates the random-phase approximation (RPA), can be satisfactorily used \cite{Likos01}. This is due to the softness of the potential which means that when the density is high, each particle interacts with many neighbours, which justifies the use of the simple mean-field approximation. In Refs.\ \cite{ARK13,ARK15} comparison between DFT and Brownian dynamics (BD) computer simulations for a different but related model 2D soft-core system of particles exhibiting freezing and also quasicrystal formation, showed good agreement between the DFT and BD results, pointing to the reliability of the DFT we use here in 2D. Also, in Ref.\ \cite{AWTK14} the present 2D GEM-4 system was studied. There it was shown that for the liquid state the RPA DFT test particle results for the radial distribution function $g(r)$ compare well with results from Hyper-netted-chain (HNC) integral equation theory \cite{HM}. These results give confidence that the simple RPA DFT is able to account well for the structure of the fluid state in 2D.

An important feature of the 2D GEM-4 system is that at temperatures above $k_BT/\varepsilon\approx0.045$ (our results here are for $k_BT/\varepsilon=0.5$), there is no hexatic \cite{Gasser09} intervening between the liquid and crystal phases \cite{PS14}. This is because the hexatic only occurs where the 2D GEM-4 fluid freezes to a single-occupancy crystal. However, at higher temperatures, the system freezes directly from the liquid to a multiple-occupancy cluster crystal \cite{PS14}. This aspect and the absence of the hexatic has implications for how our results may relate to other 2D systems. We return to this issue in our final conclusions section below.

The remainder of the paper is organised as follows: In section II we formulate the model and describe the DFT that we use in our study. Our numerical results are presented and discussed in section III. {The paper concludes with a summary and discussion in section IV}. 

\section{Model and Density Functional Theory}

We consider a 2D system of soft particles interacting with one another via a soft purely repulsive pair potential,
 \bb
 \phi(r)=\varepsilon\exp\left[-(r/R)^4\right],
 \label{phi}
 \ee
which is the GEM-4 pair potential. $r$ is the distance between the centres of the particles and the parameters $\varepsilon>0$ and $R$ define strength and range of the potential, respectively.

In DFT \cite{Evans79, Evans92, HM}, the equilibrium properties of the system are obtained by minimising of the grand potential functional
 \bb
 \Omega[\rho(\rr)]=F[\rho(\rr)]+\int\dr\rho(\rr)[V(\rr)-\mu]\,,\label{om}
 \ee
where $\rhor$ is the one-body density profile, $\mu$ is the chemical potential and $V(\rr)$ is the external potential. The intrinsic Helmholtz free energy functional $F[\rho]$ can be separated into two contributions,
 \bb
 F[\rho(\rr)]=F_{\rm id}[\rho(\rr)]+F_{\rm ex}[\rho(\rr)]\,. \label{F}
 \ee
The ideal-gas part is known exactly,
 \bb
\beta F_{\rm id}=\int\dr \rho(\rr)\left[\ln[\Lambda^2\rho(\rr)]-1\right],
 \ee
 where $\beta=1/k_BT$ is the inverse temperature, $k_B$ is Boltzmann's constant and $\Lambda$ is the thermal de Broglie wavelength. The excess part of the free energy is approximated as \cite{Likos01}
  \bb
  F_{\rm ex}[\rhor]=\frac{1}{2}\int\dr_1\int\dr_2\rho(\rr_1)\rho(\rr_2)\phi(|\rr_1-\rr_2|),
  \label{fex}
  \ee
which generates the RPA for the pair direct correlation function:
 \bb
c^{(2)}(|\rr_1-\rr_2|)\equiv-\frac{\delta^2\beta F_{\rm ex}}{\delta\rho(\rr_1)\delta\rho(\rr_2)}=-\beta v(|\rr_1-\rr_2|).
 \ee
It is well-known that the approximation in Eq.\ \eqref{fex} provides a reliable description of the thermodynamics and structure of soft particles at high densities and not too low temperatures. In this regime, the average number of neighbours per particle is large, which justifies the mean-field character of \eqref{fex}. This approximation is thus intimately connected with the ultrasoft character of the potential (\ref{phi}) and is clearly not applicable for pair potentials possessing a hard core.

Here we consider two types of external potentials, representing the influence of a solid substrate. For a planar wall, we use the repulsive Yukawa potential
 \bb
 V_p(x,z)=\left\{\begin{array}{ll} \infty\, & z<0\,,\\\label{vp}
A\frac{e^{-z/\lambda}}{z/\lambda}\,\,\,\,\,\,\,\,& z>0\end{array}\right.\,,
 \ee
where the parameter $A>0$ determines the strength and $\lambda$ the range of the wall potential, respectively. This potential approximates well the effective interaction between polymers or dendrimers with a hard wall \cite{LBHM00, BLHM01, JDLFL01, Dzubiella_2001, LBFKMH02}, as long as $\lambda\approx R_g$, where $R_g$ is the radius of gyration. We show below that the values of both $A$ and $\lambda$ are essential for determining whether or not wall induced freezing or melting occurs.

The second external potential we consider is that formed by two hard walls meeting at an opening angle $\psi$,
 \bb
 V_w(x,z)=\left\{\begin{array}{ll} \infty\;\;\;\; & {\rm for}\;z<0,\;{\rm or}\;z>\tan(\psi)x\,,\\\label{vw}
0\,& {\rm otherwise}\end{array}\right.\,.
 \ee
{The main interest in this case is to explore the influence of the value of the corner angle $\psi$ on the interfacial freezing} of the soft particles confined between these $A\to0$ walls.

For a given external potential $V(\rr)$, minimising the Grand potential (\ref{om}) leads to an Euler-Lagrange equation, which can be re-written as follows:
 \bb
\rho(\rr)= \Lambda^{-2}\exp[\beta\mu-c^{(1)}(\rr)-\beta V(\rr)]\,,\label{el}
 \ee
where the one-body direct correlation function is given by the convolution
\bb
c^{(1)}(\rr)=-\int\dr'\rho(\rr')\beta\phi(|\rr-\rr'|).
\label{eq:c_2}
\ee
Eq.\ (\ref{el}), which gives the equilibrium density distribution of the GEM-4 particles, is
solved numerically on a 2D Cartesian grid with spacing $\Delta x=\Delta z=0.05 R$. The convolution in Eq.\ \eqref{eq:c_2} is evaluated in reciprocal space with the aid of fast Fourier transforms.

\section{Results}

\begin{figure}
\includegraphics[width=0.5\textwidth]{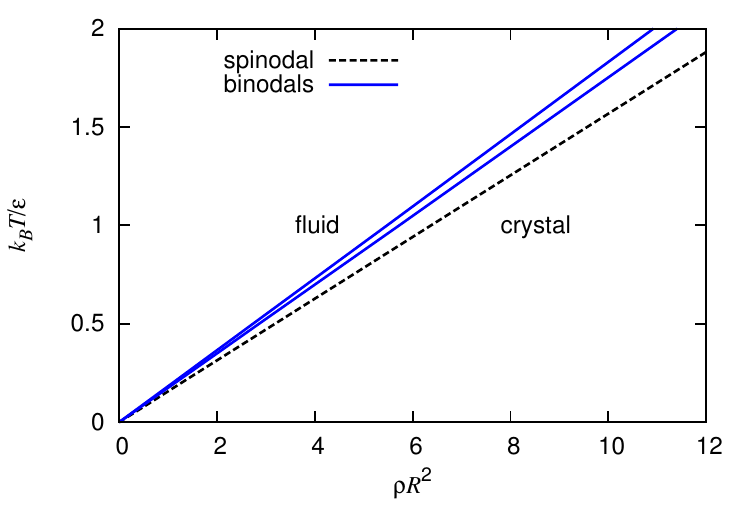}
\caption{DFT phase diagram for the 2D GEM-4 fluid. For a given value of the temperature, the binodals give the densities of the coexisting fluid and crystal phases.
We also display the `spinodal' or limit beyond which the metastable uniform liquid state is linearly unstable.} \label{pd_bulk}
\end{figure}

Before describing the properties of the 2D GEM-4 model at interfaces, we first recall the bulk phase behaviour. This was determined in Ref.\ \cite{PS14} using Monte-Carlo computer simulations. For low temperatures $k_BT/\varepsilon\lesssim0.045$, on increasing the density, the particles behave very similar to conventional 2D fluids \cite{Gasser09,KK15}, first exhibiting a transition to a hexatic phase, before freezing to form a hexagonal crystal phase. On further increasing the density, the soft-core nature becomes evident and the particles start to overlap, pairing up to form a 2-cluster crystal. Further increases in the density lead to higher numbers of particles in each cluster. See also Ref.\ \cite{WS14} for a description of the equivalent 3D behaviour. However, for temperatures $k_BT/\varepsilon\gtrsim0.045$, the fluid freezes straight to a multiple occupancy cluster crystal and there is no hexatic. For higher temperatures, this freezing transition is well described by the mean-field DFT in Eq.\ \eqref{fex} -- see \cite{PS14}. The DFT predicts a first-order freezing transition. The phase diagram is found by solving Eq.\ (\ref{el}) with $V(\rr)=0$, and is displayed in Fig.\ \ref{pd_bulk}. For a given temperature $k_BT/\varepsilon$, there is always a uniform density solution, $\rho(\rr)= \rho$, corresponding to the uniform liquid state. This profile is only a minimum free energy solution at lower densities. At higher densities, beyond the spinodal in Fig.\ \ref{pd_bulk}, it becomes a metastable saddle-point in the free energy -- i.e.\ at this point the uniform liquid state becomes linearly unstable. For higher values of the average density in the system, there is also a non-uniform density profile solution to Eq.\ (\ref{el}), corresponding to the crystal phase. For a given temperature, at a unique value of the chemical potential, these two solutions have the same value of the grand potential (i.e.\ same pressure) and thus are in thermodynamic coexistence. The average densities of these two coexisting states are the binodals displayed in Fig.\ \ref{pd_bulk}. The DFT predicts the existence of only one crystal phase. Also, since the potential (\ref{phi}) is purely repulsive, there is only one fluid state.

\begin{figure}
\includegraphics[width=0.5\textwidth]{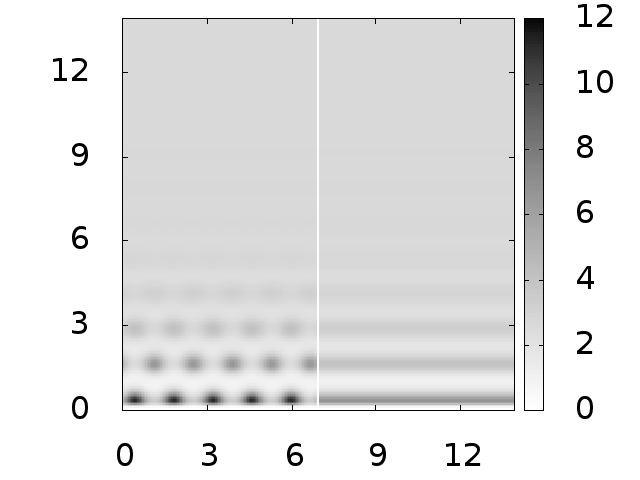}
\includegraphics[width=0.5\textwidth]{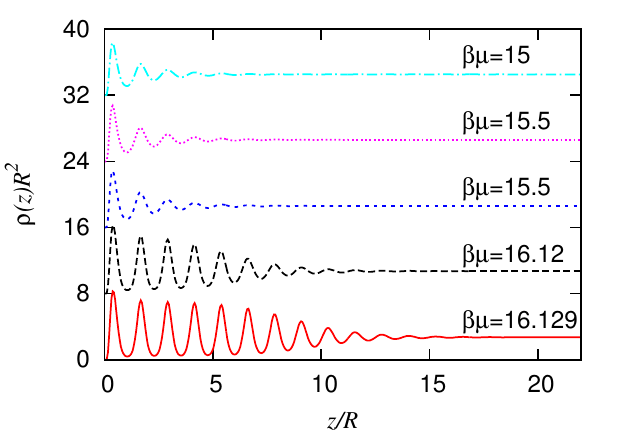}
\caption{Density profiles against a planar wall with potential \eqref{vp}, with $\beta A =1$ and $\lambda/R =1$. In the top pannel, we display the coexisting profiles corresponding to a surface freezing layer on the wall (left) and remaining fluid at the wall (right). Both profiles are for $\beta\varepsilon=2$ and $\beta\mu=15.5$, the state point at which they coexist. Below, we display a series of laterally averaged density profiles. The upper two profiles are for the liquid at the wall and the lower ones average over the density peaks corresponding to a prefrozen layer on the wall. Note that the corresponding chemical potential values are indicated and the profiles have been vertically shifted for clarity.} \label{profs_freeze}
\end{figure}

\begin{figure}
\includegraphics[width=0.5\textwidth]{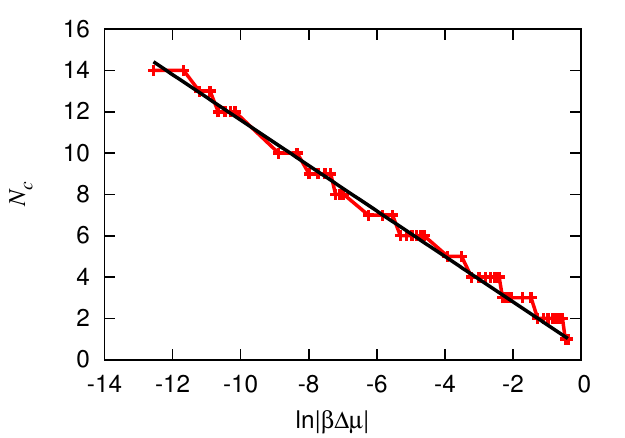}
\caption{The thickness of the surface freezing layer as coexistence is approached, for $\beta\varepsilon =2$, {$\lambda/R=1$ and $\beta A=1$. The thickness $N_c$ is defined as the number of crystalline layers at the wall. Note that $\Delta\mu\equiv \mu-\mu_{\rm coex}$, where $\mu_{\rm coex}$ is the value at coexistence. The straight line guide to the eye has gradient $m=-1.1$, indicating that the surface freezing layer thickness $N_c \sim -m\ln|\Delta\mu|$ as coexistence is approached.}} \label{peaks}
\end{figure}

{Now we describe what happens at the interface between a planar wall and the bulk fluid. The inverse temperature is set to be $\beta\varepsilon=2$, at which the system is well described by the RPA free energy functional \eqref{fex},} and the coexistence between the bulk fluid and crystal phases is predicted to occur at the chemical potential value $\beta\mu_{\rm coex}=16.1299$. In Fig.\ \ref{profs_freeze} we display density profiles for chemical potential values $\mu<\mu_{\rm coex}$, i.e., for state points where the bulk thermodynamic equilibrium phase is the fluid state. These profiles are calculated for a wall with a relatively weak repulsion strength $\beta A=1$ and range $\lambda/R =1$. In this case, the external potential does not considerably differ from the hard wall. As the chemical potential is varied, we find that when the chemical potential takes the value $\beta\mu_{\rm pf}=15.5$, where $\mu_{\rm pf}<\mu_{\rm coex}$, we observe a first-order transition from liquid ordering at the wall to crystalline ordering at the wall. {Thus, at $\mu_{\rm pf}$ a \emph{prefreezing} transition occurs. When $\mu=\mu_{\rm pf}$} two different density profiles coexist -- i.e.\ have same value of $\Omega$ -- these are displayed in the upper panel of Fig.~\ref{profs_freeze}. The density profile on the left exhibits several crystalline layers adsorbed on the surface of the wall, in contrast to the profile on the right which exhibits liquid-like structuring at the wall, which varies only in the direction perpendicular to the wall. As the chemical potential is further increased, approaching $\mu_{\rm coex}$, the number of crystalline layers adsorbed on the wall increases. This is illustrated in Fig.~\ref{peaks} where we display a plot of the number of crystalline layers adsorbed at the wall, which shows a logarithmic divergence of the total thickness of the crystalline layer as $\mu_{\rm coex}$ is approached \cite{likos}. The density profiles in the lower panel of Fig.\ \ref{profs_freeze} are obtained by first calculating the two-dimensional density profiles at the wall, such as those displayed in the upper panel of Fig.\ \ref{profs_freeze}. From these, we then average over the direction parallel to the surface of the wall to give the laterally averaged density profile:
\bb
\rho(z)=\frac{1}{L}\int_0^L\rho(x,z) dx,
\label{eq:average}
\ee
where $L$ is the width of the system.

All the observed phenomena described above appear to be analogous with the complete wetting (or drying) behaviour when a film of the liquid (gas) phase which is metastable in bulk is formed at the wall-gas (wall-liquid) interface \cite{dietrich,evans90}. Recall that for wetting, if the transition (at bulk coexistence) is first order at the wetting temperature $T_w$, there is indeed an off-coexistence extension of the free energy singularity, giving rise to a  first-order prewetting or thin-to-thick adsorbed film transition. Also, when the interactions between the particles are short ranged, then (above $T_w$) the thickness of the adsorbed layer, $l\sim-\ln|\Delta\mu|$, where $\Delta\mu\equiv\mu-\mu_{\rm coex}$. This is similar to the prefreezing we observe here. However, there is an important difference between wetting and wall induced freezing. In the wetting (or drying) case, one observes a preferential adsorption of a phase possessing the \emph{same} symmetry as the bulk phase, while in the prefreezing case the symmetry between the two coexisting phases is \emph{different}. The change in the translation symmetry (from continuous to discrete) thus necessitates that the occurrence of the crystalline phase to be a first order transition. It is interesting to note that it is the symmetry breaking along the direction parallel to the wall which drives the prefreezing transition, and there is almost no change in the excess number of particles adsorbed at the wall as $\mu_{\rm pf}$ is crossed. This can be seen from the lower panel of Fig.~\ref{profs_freeze}. For {$\beta\mu=\beta\mu_{\rm pf}=15.5$}, comparing the laterally averaged density profiles of the coexisting states displayed in the upper panel, we see they are almost indistinguishable.

\begin{figure}
\includegraphics[width=0.5\textwidth]{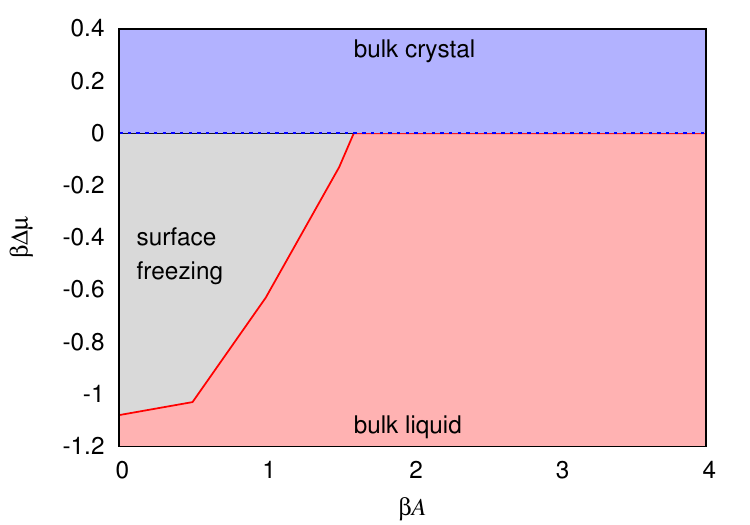}
\caption{The interfacial phase diagram as a function of $A$, for $\lambda/R = 1$ and $\beta\varepsilon = 2$. The limit $A = 0$ corresponds to the hard wall, at which prefreezing occurs. As $A$ is increased, the interval in $\Delta\mu$ in which prefreezing occurs gets smaller, and for $\beta A\ge 1.6$ there is no prefreezing at all.} \label{spd_l=1}
\end{figure}

In Fig.~\ref{spd_l=1} we display the surface phase diagram, showing the locus of the prefreezing transition as the wall repulsion strength parameter $A$ is varied, while the range of the wall potential is kept fixed at $\lambda/R=1$. When $A<A_c$, where $\beta A_c\approx1.6$, the wall induces freezing, accompanied by a first-order prefreezing transition. When $A>A_c$, no surface freezing occurs and in the limit $\Delta\mu\equiv(\mu-\mu_{\rm coex})\to0^-$, the whole system is filled with the (strongly inhomogeneous) liquid phase. For $\Delta\mu\to0^+$, the whole system remains in the crystal state for all values of $A$. This phenomenology can be easily understood by recalling that all interactions in the system are purely repulsive. When the strength of the wall repulsion is less than the particle-particle interaction, there is a strong affinity of the particles to the wall surface, because this reduces the total energy of the system. This can be most clearly understood in the $A=0$ limit, when the external potential acts as a hard-wall. In this limit, a particle sitting on the surface of the wall interacts with only a half-space containing particles, so the potential energy is less than the value for a particle in the bulk, far from the wall. The effect of this is that when $A$ is small, the density of particles on the surface of the wall is high. If the surface density is high enough, then these particles can further lower the energy by fully overlapping with some particles, so as to decrease the overlaps with others -- i.e.\ the surface particles freeze into a cluster-crystalline layer. On increasing $A$, this effect becomes weaker, as does the corresponding transition, which eventually disappears at $A_c$. Since the freezing is always first-order, there is no analog to the surface critical point of the prewetting transition.

\begin{figure}
\includegraphics[width=0.5\textwidth]{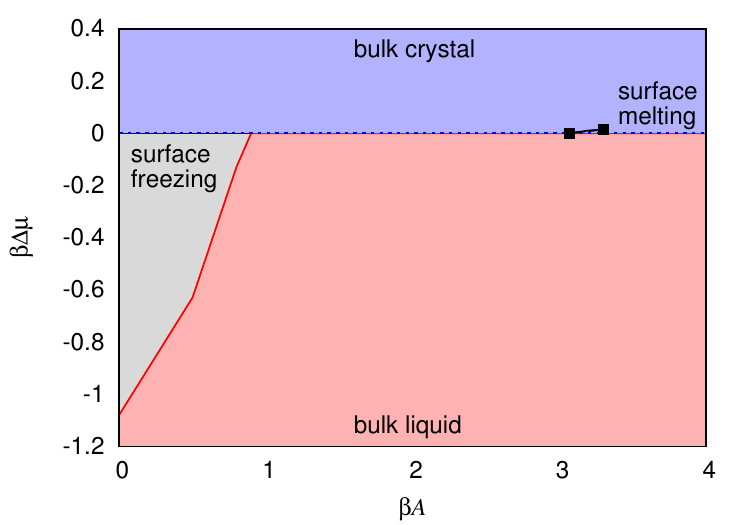}
\caption{The interfacial phase diagram as a function of $A$, for $\lambda/R =4$ and $\beta\varepsilon =2$. The limit $A =0$ corresponds to the hard wall, at which surface freezing occurs. As $A$ is increased, the interval in $\Delta\mu$ in which surface freezing (or prefreezing) occurs gets smaller, and for $\beta A\ge0.9$ there is no surface freezing when the bulk is the liquid phase. However, in contrast to the case when $\lambda/R =1$, on approaching coexistence from above (when the crystal is the bulk phase), we find that for $\beta A\ge3.07$ there is wetting of the wall-crystal interface by the liquid phase. This wetting transition is first order, with the usual prewetting line. However, the prewetting line is rather short -- in the figure above the ends of the prewetting line are marked with ``$\blacksquare$''.} \label{spd_l=4}
\end{figure}

\begin{figure}
\includegraphics[width=0.5\textwidth]{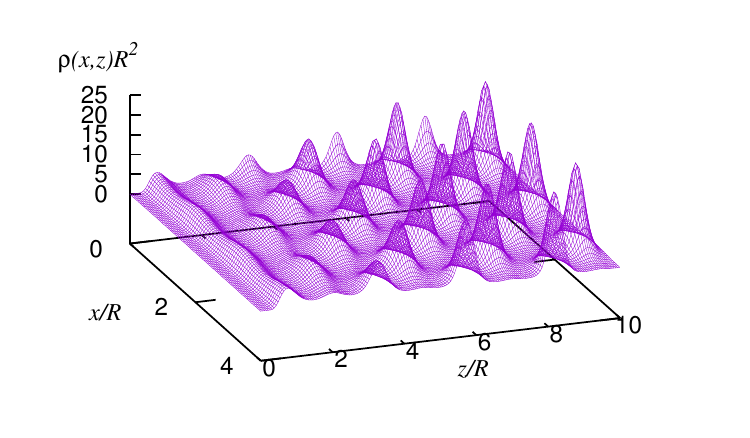}
\caption{2D density profile of the crystal phase near the wall with $\beta A=4$ and $\lambda/R=4$. The surface of the wall is along the line $z=0$. The chemical potential $\beta\mu=16.25$ and temperature $k_BT/\varepsilon=0.5$. Note that there is surface melting, i.e.\ the density profile near the wall is liquid-like. However, right at the wall there are still small amplitude oscillations in the profile parallel to the wall due to the proximity of the bulk crystal.} \label{70}
\end{figure}

\begin{figure}
\includegraphics[width=0.5\textwidth]{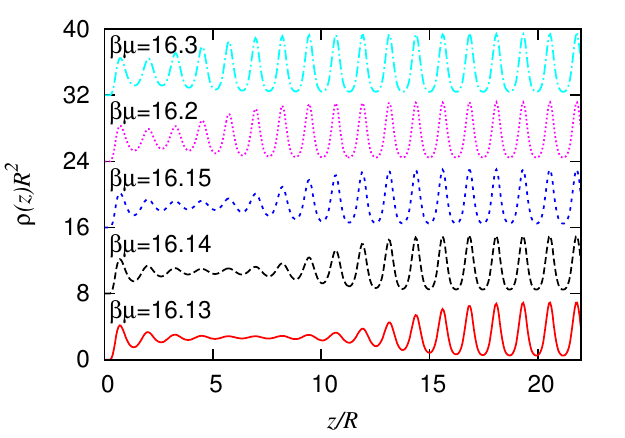}
\caption{A series of laterally averaged density profiles against a planar wall with $\beta A = 3.7$ and $\lambda/R=4$. The upper profiles, which are for state points away from coexistence, show the crystal right up to the wall, but as coexistence at $\beta\mu_{\rm coex} =16.1299$ is approached, we see in the lower profiles that there is a growing liquid film on the wall. This film of liquid grows continuously in thickness as coexistence is approached, with diverging thickness at coexistence. Note that the corresponding chemical potential values are indicated and the profiles have been vertically shifted for clarity.} \label{profs_A=3.7}
\end{figure}

\begin{figure}
\includegraphics[width=0.5\textwidth]{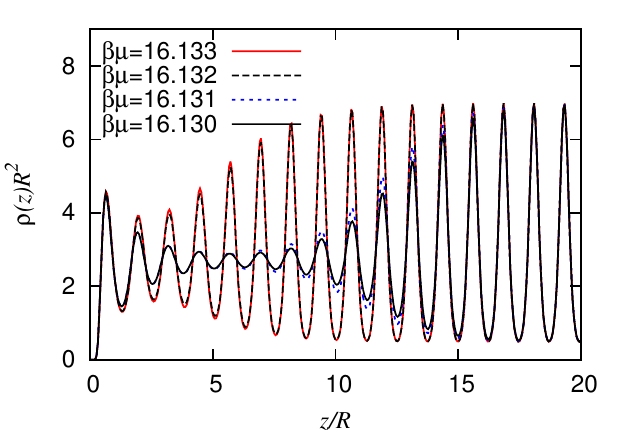}
\caption{A series of laterally averaged density profiles in the vicinity of a planar wall with $\beta A =3.1$ and $\lambda/R=4$. {Profiles for state points at either side of the prewetting (or surface melting) line are displayed}. For chemical potential values $\beta\mu\geq16.132$ we observe the crystal phase right up to the wall, but for $\beta\mu<16.132$ we observe a layer of the liquid phase adsorbed at the wall. The thickness of this liquid layer then increases as coexistence is approached, diverging as $\mu\to\mu_{\rm coex}^+$, where $\beta\mu_{\rm coex} =16.1299$.} \label{profs_A=3.1}
\end{figure}

\begin{figure*}

\includegraphics[width=0.32\textwidth]{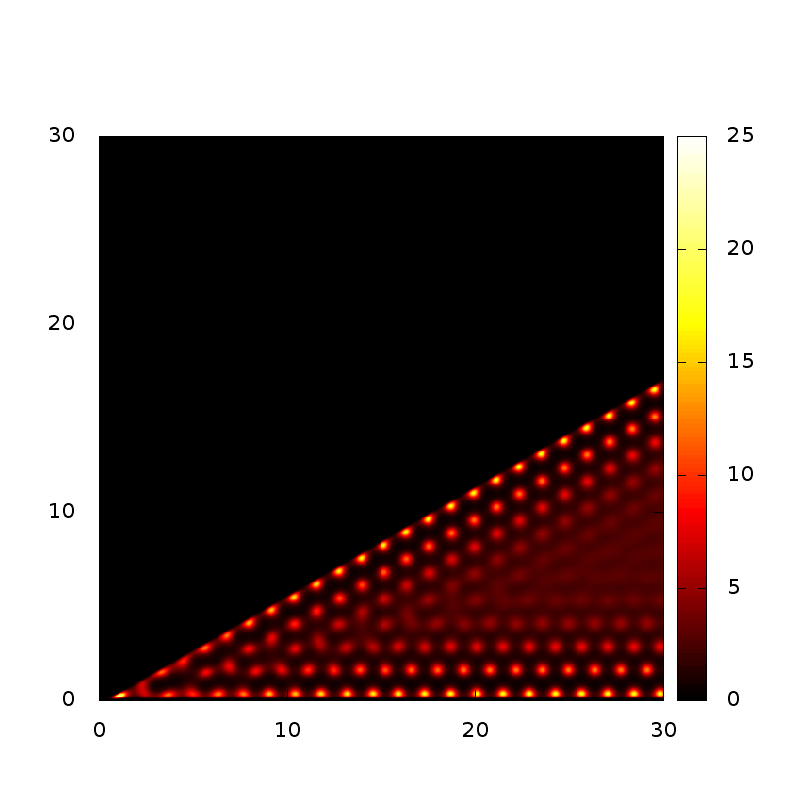}
\includegraphics[width=0.32\textwidth]{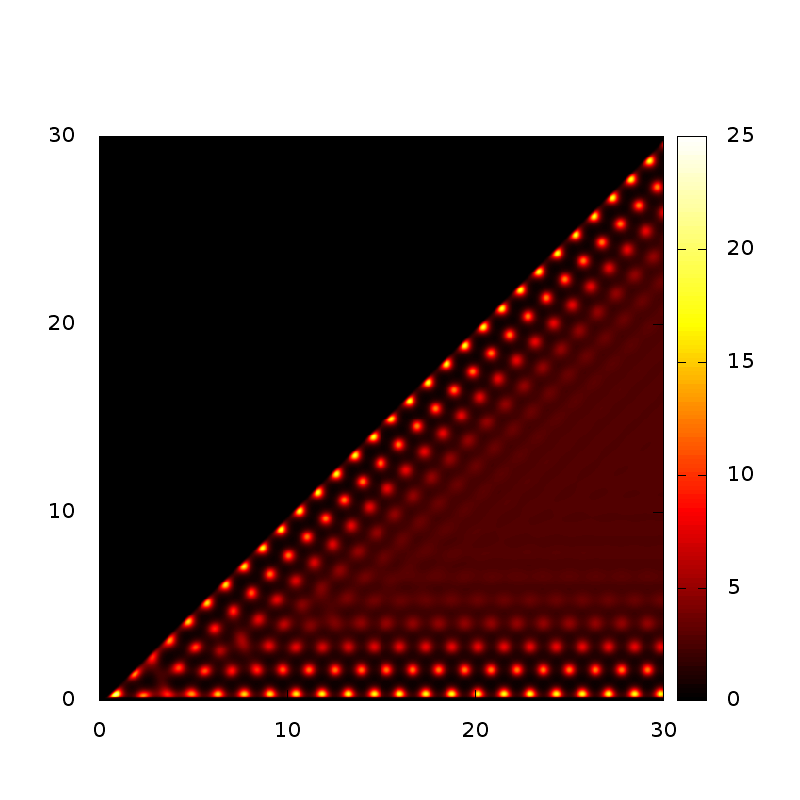}
\includegraphics[width=0.32\textwidth]{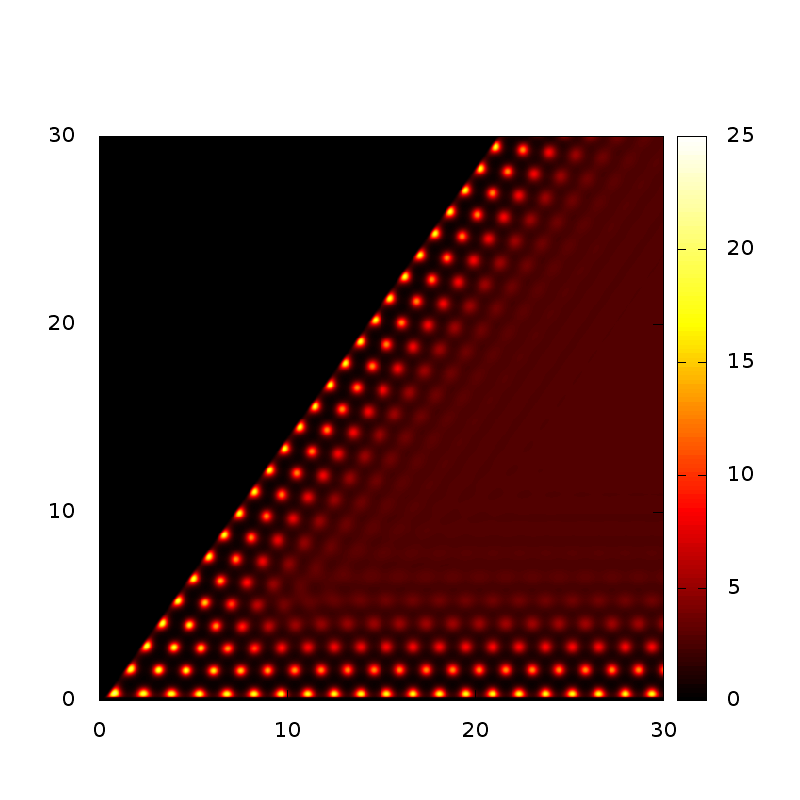}

\includegraphics[width=0.32\textwidth]{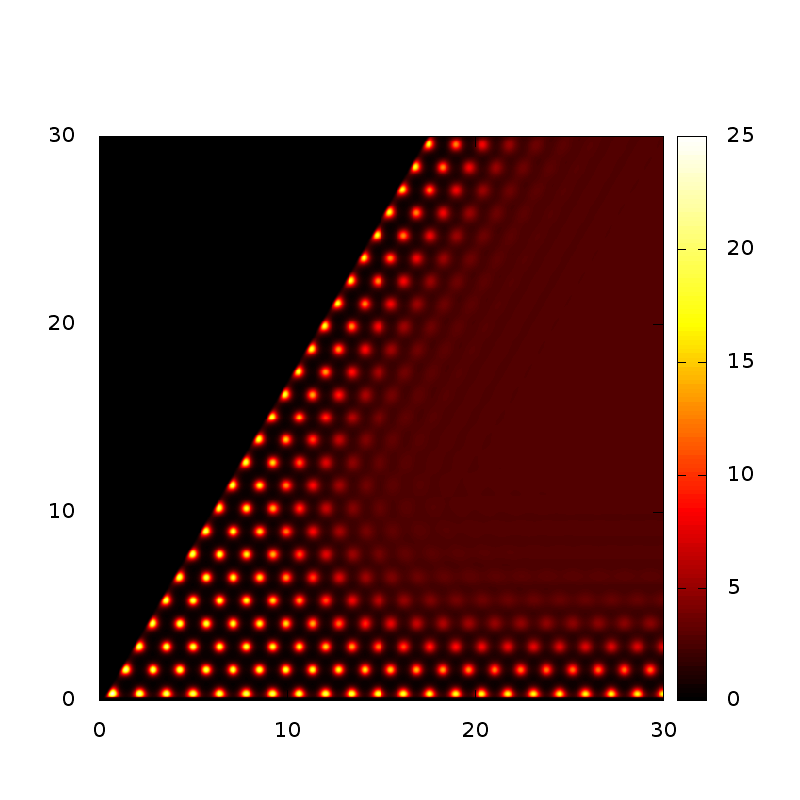}
\includegraphics[width=0.32\textwidth]{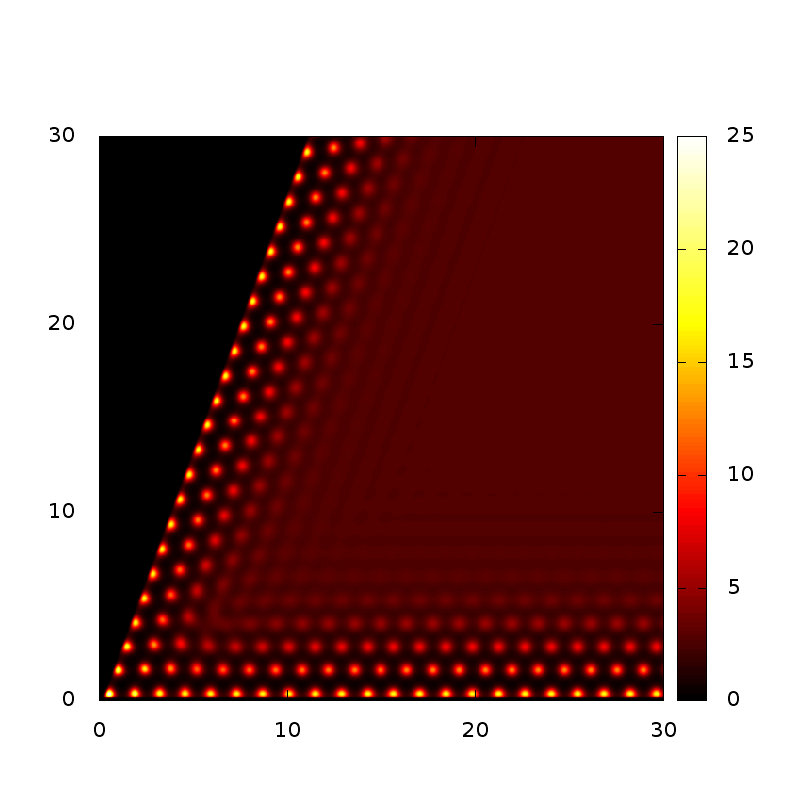}
\includegraphics[width=0.32\textwidth]{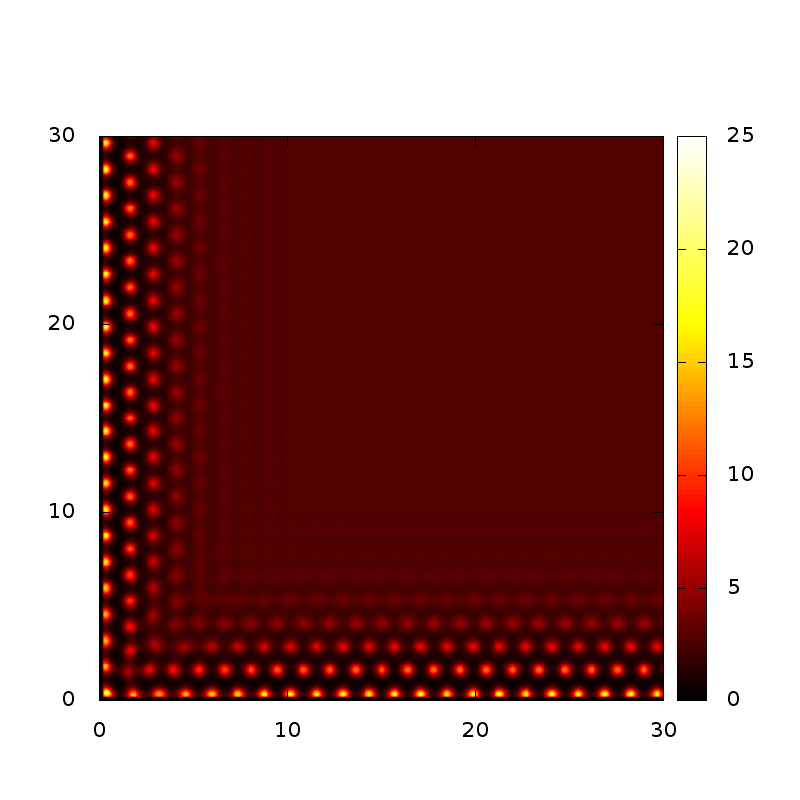}
\caption{Density profiles of the fluid confined in a hard-wall wedge with varying opening angle, $\psi$. From top left to bottom right: $\psi= 30\degree, 45\degree, 55\degree, 60\degree, 70\degree, 90\degree$. Note that the particles are frozen in the vicinity of the wall. Depending on $\psi$, the corner contains more or less frozen layers than at the planar wall. The chemical potential $\beta\mu=15.8$, which corresponds to the fluid state in bulk. }\label{profs_wedge}
\end{figure*}

\begin{figure}
\includegraphics[width=0.5\textwidth]{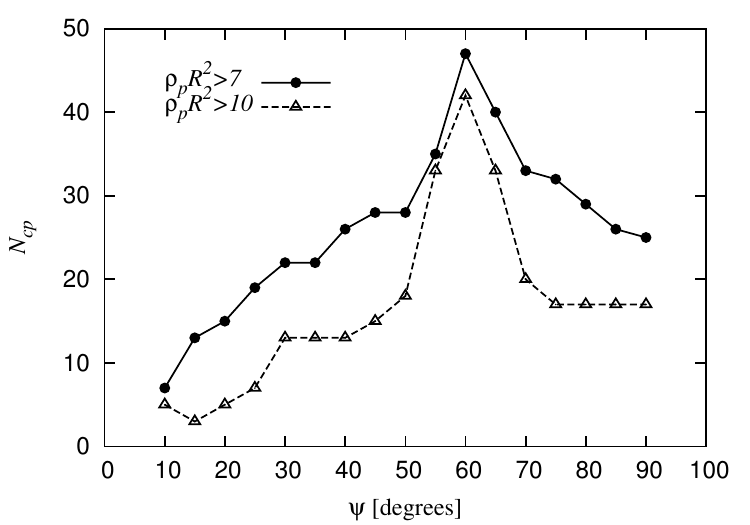}
\caption{Number of crystalline density peaks in the corner $N_{\rm cp}$, defined as the number of density peaks within a distance of $15R$ from the corner of the wedge. We plot the number of peaks with a peak density value $\rho_{\rm p}R^2>7$ (solid line) and $\rho_{\rm p}R^2>10$ (dashed line).} \label{peaks_apex}
\end{figure}

In Fig.\ \ref{spd_l=4} we display the interfacial phase diagram for the case when the wall potential range is increased to $\lambda/R=4$. Compared to the previous case in Fig.\ \ref{spd_l=1}, with $\lambda/R=1$, the net wall repulsion is greater for a given $A$ and thus the prefreezing line is steeper and meets the bulk coexistence line at $\beta A\approx0.9$, rather than at $\beta A\approx1.6$ (compare Figs.\ \ref{spd_l=1} and \ref{spd_l=4}). For $\beta A>0.9$ the slowly decaying wall repulsion prevents prefreezing and hinders the formation of the crystalline structure near the wall surface. Furthermore, in contrast with the case with $\lambda/R=1$, when the bulk phase is the crystal, the wall now has the ability to induce \emph{surface melting} when $A>A_{\rm m}$, where $\beta A_{\rm m}=3.07$. This is the inverse process to surface freezing, occurring as coexistence is approached from above, $\Delta\mu\to0^+$. When the wall repulsion is sufficiently strong (relative to the particle-particle repulsion), the slowly decaying wall potential decreases the density near the surface of the wall, which gives rise to a liquid-like structure in the vicinity of the wall. In Fig.\ \ref{70} we display the density profile at the wall-crystal interface when the chemical potential $\beta\mu=16.25$, which corresponds to $\beta\Delta\mu\approx0.12$, where the average bulk crystal density $\bar{\rho}R^2=2.8$. The density profile is liquid-like at the surface of the wall, i.e.\ surface melting occurs. However, one can also see small amplitude density modulations in the liquid at the wall due to the vicinity to the bulk crystal so that the liquid phase still exhibits broken translation symmetry along the $x$-axis. This is in contrast to the surface freezing, where the formation of the new crystalline phase requires a change of symmetry, whereas for surface melting the symmetry of both phases is the same, albeit the amplitude of the density modulations in the liquid can be small. The consequence of this is that it allows for the existence of a critical point at $A=A_{\rm mc}$, separating a regime where the surface melting appears via a first order transition as $\Delta\mu\to0^+$, from a regime where the thickness of the liquid film at the interface grows continuously. Therefore, the topology of the surface phase diagram for premelting is the same as for wetting/drying.

In Fig.~\ref{profs_A=3.7} we display a series of laterally averaged density profiles for state points approaching coexistence $\Delta\mu\to0^+$, for the wall with $\lambda/R=4$ and $\beta A=3.7$. Laterally averaging over the density profile of the crystal phase using Eq.\ \eqref{eq:average} results in a highly oscillatory profile; each oscillation corresponds to a different crystal plane. The results in Fig.~\ref{profs_A=3.7} are typical of the case $A>A_{\rm mc}$. The amplitude of the oscillations are diminished close to the wall. For states near to coexistence, the profiles exhibit a portion where the density is almost flat, corresponding to a film of the liquid. The thickness of the liquid layer $l$ grows as coexistence is approached, diverging at coexistence $l\sim-\ln|\Delta\mu|$. For $A>A_{\rm mc}$ and for any value of the chemical potential there is a unique minimum solution of the grand potential functional (\ref{om}). Thus, the surface melting is continuous. In contrast, when the wall strength $A_{\rm m}<A<A_{\rm mc}$ there are two solutions to Eq.~(\ref{om}): one corresponding to crystal right up to the wall and the other to a state exhibiting a liquid-like layer adsorbed at the wall. However, this fluid layer still possesses lateral inhomogeneity. The line at which the two branches of solutions intersect in the $\Omega$-$\mu$ plane defines the first-order \emph{premelting} transition. In Fig.~\ref{profs_A=3.1} we display a series of laterally averaged density profiles, calculated along a path approaching coexistence. As the premelting line is crossed, the liquid-like layer forms discontinuously, appearing in Fig.~\ref{profs_A=3.1} as a sudden jump in the amplitude of the oscillations near the wall. The thickness of the liquid layer eventually diverges as $\Delta\mu\to0^+$.

\begin{figure}
\includegraphics[width=0.5\textwidth]{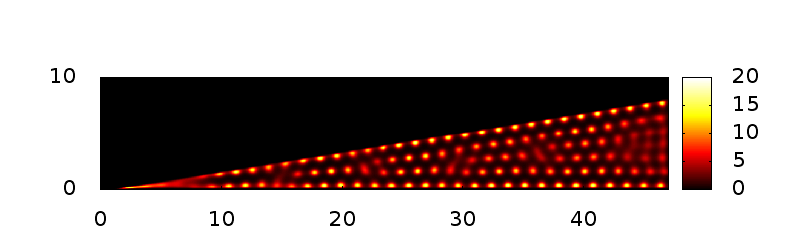}
\caption{Density profile in a hard-wall wedge with an opening angle $\psi=10\degree$. The chemical potential $\beta\mu=15.8$ corresponds to the fluid state in
bulk.}\label{prof_wedge10}
\end{figure}

{Now consider} the crystallization of the GEM-4 fluid inside a linear wedge formed by the conjunction of two hard walls. Our primary interest is in the influence on the surface freezing of the opening angle $\psi$ of the corner at which the two walls meet. In Fig.~\ref{profs_wedge} we display a series of equilibrium density profiles for $\beta\Delta\mu=-0.33$ and for various values of $\psi$. As discussed above, at the planar hard-wall prefreezing occurs when the bulk fluid state is near to coexistence with the crystal. For this state point, far from the apex of the wedge, this surface frozen layer consists of just two pronounced crystalline layers. Near the corner, the degree to which the crystal structure can be accommodated between the two converging walls has a significant effect on the amount of frozen material near the apex where the two crystalline films meet. {When} $\psi=60\degree$, the thickness of the crystalline layer in the corner is much thicker than on the planar wall and also for any of the other cases in Fig.\ \ref{profs_wedge}. For the case when $\psi=55\degree$, which is close to $\psi=60\degree$, {there is} a thickening of the crystalline layer, but to a much lesser extent. This is due to the fact that when $\psi=60\degree$, the two walls are {\em both} parallel to crystal planes and so the effect of the corner is to enhance the amount of frozen particles in the vicinity of the corner.

\begin{figure*}
\includegraphics[width=0.32\textwidth]{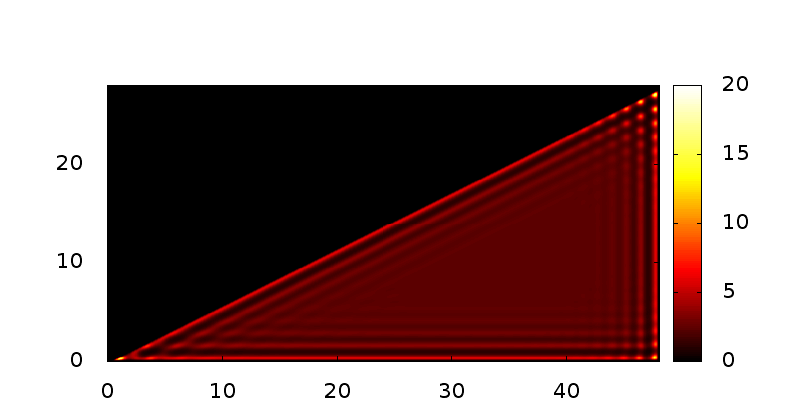}
\includegraphics[width=0.32\textwidth]{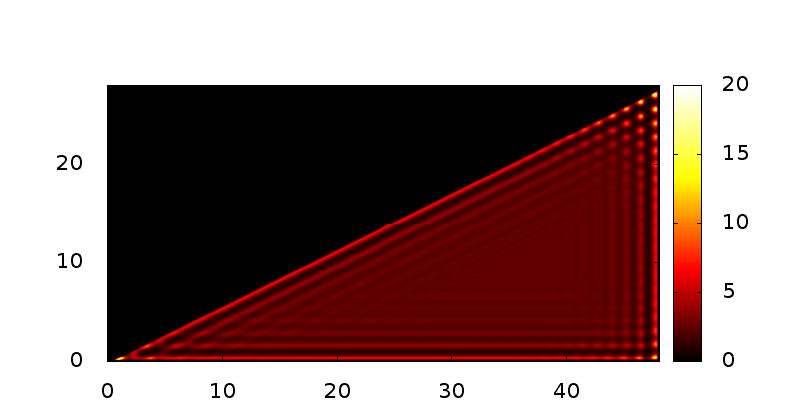}
\includegraphics[width=0.32\textwidth]{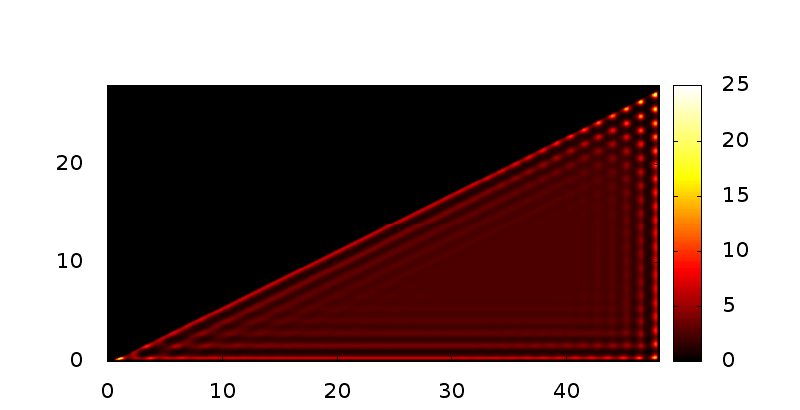}

\includegraphics[width=0.32\textwidth]{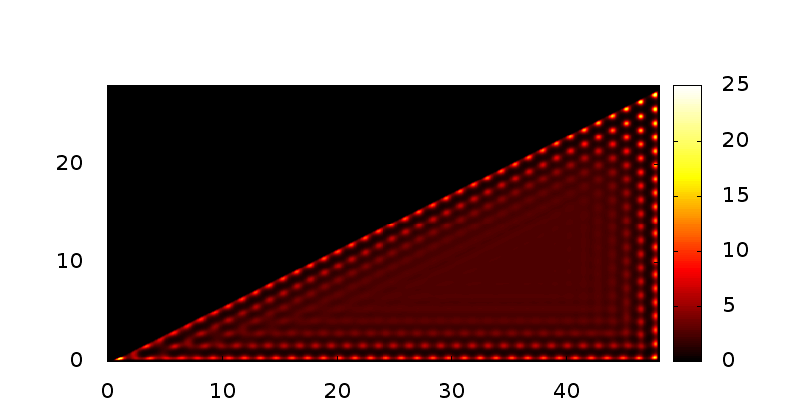}
\includegraphics[width=0.32\textwidth]{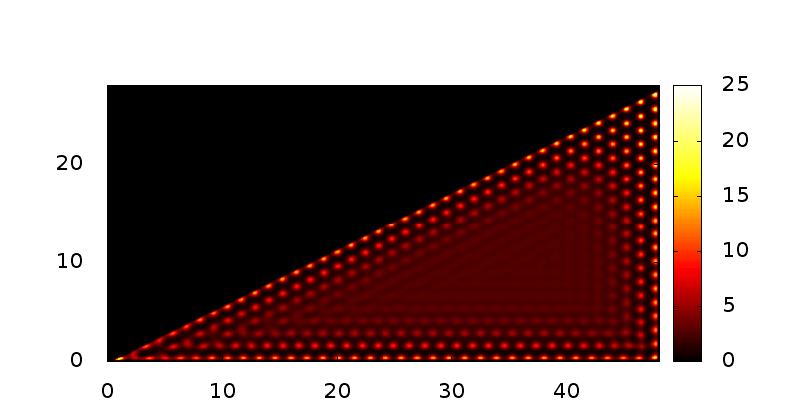}
\includegraphics[width=0.32\textwidth]{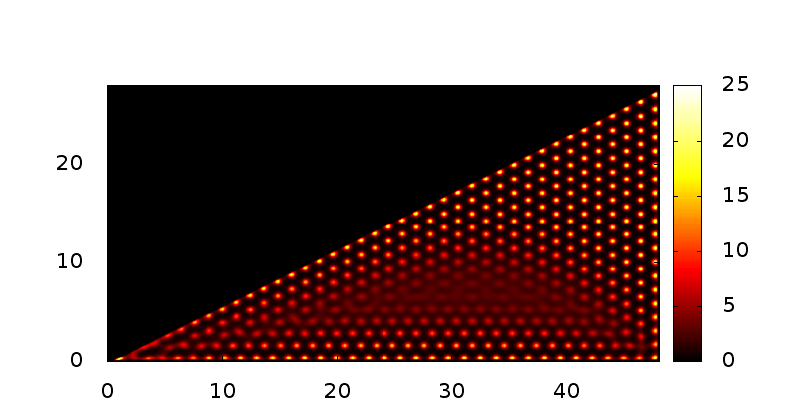}
\caption{Density profiles for the fluid confined within a triangle, as the chemical potential is varied, approaching bulk coexistence. The chemical potential values, going top left to bottom right, are: $\beta\mu =14.5, 15, 15.25, 15.5, 15.75$ and $16$. The walls of the triangle are hard and the three corner angles are $30\degree$, $60\degree$ and $90\degree$.} \label{trian}
\end{figure*}

In Fig.\ \ref{peaks_apex} we display a plot of $N_{\rm cp}$, the number of sharp (i.e.\ crystal-like) density peaks in the corner, as a function of the corner angle $\psi$. $N_{\rm cp}$ is defined as the number of density peaks with a peak value $\rho_{\rm p}$ greater than a specified threshold value that are located at a distance less than $15R$ of the apex of the corner. We display $N_{\rm cp}$ for peaks with $\rho_{\rm p}R^2>7$ and also with $\rho_{\rm p}R^2>10$. Qualitatively, the results are very similar over the interval $\psi\in(10\degree,90\degree)$ with either threshold value. As $\psi$ is increased, we observe an increase in $N_{\rm cp}$, reflecting simply the fact that as the angle $\psi$ increases, the corner gets bigger and so there is more space for density peaks to be present. However, for a narrow interval of angles around the value $\psi=60\degree$, we see a significant increase in $N_{\rm cp}$, reflecting the fact that for this value the crystal phase fills the wedge to a large extent, as can be seen in in Fig.~\ref{profs_wedge}. The very strong enhancement of the crystalline structure when the opening angle $\psi\approx60\degree$ is because the wedge geometry matches with the hexagonal lattice of the 2D crystal and therefore promotes the crystallization. On the other hand, for smaller corner angles, such as when $\psi=30\degree$, there is a strong interference between the two surface layers, which suppresses the formation of the crystal near the apex. However, as can be seen from Fig.\ \ref{prof_wedge10}, for extremely acute wedges $\psi\approx10\degree$, formation of bridges connecting crystal-layers on the respective sides of the wedge occurs (cf. Fig.~\ref{peaks_apex}).

Finally, in Fig.~\ref{trian} we show a sequence of density profiles for the system confined within a triangular cavity with hard walls meeting at corners with angles $\psi=30\degree$, $60\degree$ and $90\degree$. The profiles are for a sequence of increasing chemical potential values $\mu\to\mu_{\rm coex}^-$. Placing the three different corners within the same system allows one to see the difference in how the crystal phase appears at a different rate depending on the angle of the corner as $\mu\to\mu_{\rm coex}^-$. As already observed, in the $60\degree$ corner the crystal is strongly enhanced in comparison with the other two corners. A specific feature of this closed system is that it is rather resistant to being completely frozen, even when $\mu\approx\mu_{\rm coex}$. This is because each of the corners generates a different crystal orientation and complete freezing requires the crystalline orientation induced at one of the corners to prevail over the other two. Therefore, the unfavourable crystalline structure near the $30\degree$ and $90\degree$ corners creates a barrier to full crystallization, in contrast with an open wedge.

\section{Summary and Conclusions}

Using DFT we have studied the surface freezing and surface melting interfacial phase behaviour of a soft matter system at a planar wall and inside of wedges at state points close to phase coexistence between the liquid and the {crystal, which occurs when $\mu=\mu_{\rm coex}$. For the planar wall, both the range and the strength of the repulsion due to the wall can have a profound effect on the nature of the surface phase behaviour. When} the wall repulsion is strong enough and slowly decaying, then this favours premelting when $(\mu-\mu_{\rm coex})\to 0^+$, but when the wall is repulsion decreases rapidly, then this favours prefreezing when $(\mu-\mu_{\rm coex})\to 0^-$.

In general, there is a threshold value of the wall repulsion amplitude $A_c$ below which the wall induces freezing at $\mu_{\rm coex}$ and there is a prefreezing transition on the approach to coexistence $(\mu-\mu_{\rm coex})\to 0^-$, at which the first crystalline layers near the wall nucleate. Further approaching bulk coexistence, the number of crystalline layers $N_c$ adsorbed at the wall increases and eventually diverges, {$N_c\sim-m\ln|\mu-\mu_{\rm coex}|$. This implies that the leading order term in the binding potential in an effective interfacial Hamiltonian description of surface freezing $\sim e^{-N_c/m}$, which is the dominant term as $N_c\to\infty$.} Since the prefreezing requires a change of symmetry, the transition is always first-order and cannot exhibit a critical point. This feature is analogous to a bulk solid-fluid transition and distinguishes surface freezing from the wetting/drying transition, where the two phases have the same symmetry.

When the wall potential is slowly decaying, then there is the possibility of surface melting for sufficiently large values of the wall repulsion amplitude $A$. In this case, the wall is completely wet by the fluid as bulk phase coexistence is approached $(\mu-\mu_{\rm coex})\to 0^+$. {It is} also observed that the surface melting is not merely the inverse process to surface freezing. While in the case of surface freezing the bulk liquid possesses a different symmetry to that of the wall-adsorbed crystal, in the case of surface melting the wall-adsorbed fluid state still exhibits a broken symmetry due to its proximity to the bulk crystal. The latter are distinguished by the magnitude of the density oscillations in each phase. This implies the possibility of a critical point for the premelting transition, separating the regime for which the melting is continuous from where it is a first-order transition.

When the fluid is confined within a wedge made of two converging hard walls, the value of the opening angle of the wedge is found to have a considerable effect on the structure of the confined particles in the wedge. Whether or not the crystal structure is commensurate with the wedge shape can either promote or suppress the fluid from freezing. In particular, when the opening angle is around $\psi=60\degree$, the wedge geometry matches with the bulk crystal lattice, which strongly enhances freezing near the apex. This effect was also observed in studies of heterogeneous crystal nucleation \cite{sear, sear3}, where the angle of the groove in the 3D surface was observed to have a significant effect on the nucleation {rate.}

In Ref.\ \cite{likos} the phase behaviour of the 3D GEM-8 system confined within a slit was studied. The authors considered confinement between a pair of parallel planar walls with potentials given by the same repulsive Yukawa form as in Eq.\ \eqref{vp}, with $\lambda/R=1$ and $\beta A=10$, and also between a pair of attractive Lennard-Jones type walls. When the slit width is large, surface melting was observed at the repulsive wall and surface freezing at the attractive wall. However, in light of the present work one should also expect to observe surface freezing of the 3D GEM-8 fluid at the repulsive Yukawa wall if the range of the wall repulsion is decreased and/or the amplitude of the repulsion $A$ is decreased. The general trends and overall qualitative behaviour of the 2D GEM-4 fluid studied here should apply more generally to 2D and 3D soft-core fluids exhibiting freezing to a cluster crystal.

We should caution against drawing conclusions from our results here regarding the interfacial phase behaviour of the 2D GEM-4 fluid to 2D fluids in general, particularly those composed of particles having a hard core. Recall that in 2D, in view of the Mermin-Wagner theorem (see \cite{Gasser09} and references therein), one should expect large-scale fluctuations to prevent genuine long-range translational order. Additionally, as the density is increased, 2D fluids composed of particles with a hard-core first exhibit a transition to the hexatic phase, before freezing to form a hexagonal crystal phase \cite{Gasser09,KK15}. This is also what the 2D GEM-4 system does at low temperatures. However, for higher temperatures $k_BT/\varepsilon\gtrsim0.045$ the 2D GEM-4 system freezes from the liquid state directly to a cluster crystal state and there is no hexatic \cite{PS14}. It is in this regime that our results are relevant.

A further aspect that we should remark on relates to the mean-field character of the DFT that we use. The theory is remarkably accurate for describing the 2D GEM-4 system, due to the softness of the particles. At high densities, there are multiple overlaps and each particle interacts with very many neighbours (the classic mean-field scenario). Nonetheless, we expect fluctuation effects to round the observed surface transitions and prevent them from being genuine phase transitions. In view of the pseudo one-dimensional character of the premelting critical point, going beyond mean-field and including a proper description of the large-scale fluctuations will destroy the observed premelting transition, making the transition rounded -- see also Ref.\ \cite{Presti1D}. However, this should only apply for the 2D system studied here. We do expect a true premelting transition to occur for the 3D fluid. As regards the influence of fluctuations on the observed prefreezing transition, it is hard to judge. Even in bulk, the true 2D crystal only has quasi-long-range positional order due to fluctuations, so one must expect fluctuation effects to change the character of the transition. However, since the prefreezing requires a change in symmetry and possesses no critical point, fluctuation effects may be less pronounced than for the premelting transition. We certainly expect that in 3D the premelting transition to be qualitatively similar to that observed here based on a mean-field treatment of the 2D system.

\begin{acknowledgments}
 \noindent AM acknowledges a support from the Czech Science Foundation, project 13-09914S.
\end{acknowledgments}

\end{document}